\documentclass[times, twoside]{StyleArXiv}
\usepackage{booktabs}
\usepackage{multicol,multirow}
\usepackage{orcidlink}
\usepackage{subcaption}
\usepackage{cleveref}
\crefname{figure}{Fig}{Figs}
\crefname{table}{Tab}{Tabs}
\crefname{equation}{eq.}{eqs.}

\leadauthor{Siddhad} 

\begin{document}

\title{DrowzEE-G-Mamba: Leveraging EEG and State Space Models for Driver Drowsiness Detection}
\shorttitle{DrowzEE-G-Mamba: Leveraging EEG and State Space Models for Driver Drowsiness Detection}

\author[1,\Letter]{Gourav~Siddhad~\orcidlink{0000-0001-5883-3863}}
\author[1]{Sayantan~Dey~\orcidlink{0009-0009-4210-8435}}
\author[1]{Partha~Pratim~Roy~\orcidlink{0000-0002-5735-5254}}
\affil[1]{Department of Computer Science and Engineering, Indian Institute of Technology, Roorkee, Uttarakhand, 247667, India}

\maketitle


\begin{abstract}
Driver drowsiness is identified as a critical factor in road accidents, necessitating robust detection systems to enhance road safety. This study proposes a driver drowsiness detection system, DrowzEE-G-Mamba, that combines Electroencephalography (EEG) with State Space Models (SSMs). EEG data, known for its sensitivity to alertness, is used to model driver state transitions between alert and drowsy. Compared to traditional methods, DrowzEE-G-Mamba achieves significantly improved detection rates and reduced false positives. Notably, it achieves a peak accuracy of 83.24\% on the SEED-VIG dataset, surpassing existing techniques. The system maintains high accuracy across varying complexities, making it suitable for real-time applications with limited resources. This robustness is attributed to the combination of channel-split, channel-concatenation, and channel-shuffle operations within the architecture, optimizing information flow from EEG data. Additionally, the integration of convolutional layers and SSMs facilitates comprehensive analysis, capturing both local features and long-range dependencies in the EEG signals. These findings suggest the potential of DrowzEE-G-Mamba for enhancing road safety through accurate drowsiness detection. It also paves the way for developing powerful SSM-based AI algorithms in Brain-Computer Interface applications.
\end{abstract}


\begin{corrauthor}
g\_siddhad\at cs.iitr.ac.in
\end{corrauthor}


\section{Introduction}
\label{sec_intro}

Driver drowsiness detection is crucial for road safety, as fatigue and sleepiness are major causes of car crashes, often leading to severe injuries or fatalities. Unlike intoxication, drowsiness develops gradually and can be unnoticed by drivers. Effective detection systems can prevent accidents by alerting drivers to take corrective actions, such as resting. With the rise of advanced driver-assistance systems (ADAS)~\cite{nidamanuri2021progressive} and autonomous vehicles, integrating robust drowsiness detection is essential for enhancing transportation safety and reliability. These systems not only protect individual drivers but also contribute to public safety by reducing drowsiness-induced accidents.

EEG is a valuable tool for real-time detection and analysis of cognitive states, capturing the brain’s electrical activity~\cite{siddhad2024enhancing}. EEG measures voltage fluctuations from neuronal ionic currents, offering insights into mental states like attention, alertness, fatigue, and cognitive load~\cite{panwar2024eeg}. Its high temporal resolution is ideal for monitoring rapid changes in brain activity, making it perfect for transient cognitive state monitoring. By analyzing frequency bands (delta, theta, alpha, beta, and gamma) and spatial distribution, researchers can infer neural mechanisms behind various cognitive processes~\cite{newson2019eeg}. This capability is crucial in brain-computer interfaces (BCIs), neurofeedback, and cognitive neuroscience research. EEG’s non-invasive nature and relatively low cost enhance its practicality for cognitive state detection, advancing both clinical and real-world applications.

Due to the complex, non-linear nature of EEG data, standard deep learning models struggle with accurate analysis. This study explores Mamba~\cite{gu2021combining}, a state-of-the-art state-space model (SSM), for effective driver drowsiness detection using EEG signals. Mamba excels at capturing the intricate patterns and non-linearities within EEG data. It extracts relevant features and integrates them with a hidden state space, reflecting the underlying brain activity. This allows Mamba to effectively manage noise and uncertainties inherent in EEG data, leading to more accurate drowsiness detection. Additionally, Mamba's efficient feature extraction and adaptive learning capabilities make it ideal for real-time monitoring and prediction, surpassing traditional EEG-based methods. Building upon the advantages of structured SSMs~\cite{gu2021combining}, Mamba offers computational efficiency and excels at capturing long-range dependencies within data. Notably, Mamba addresses limitations of previous models by incorporating time-varying parameters and employing a novel hardware-aware algorithm for efficient training and inference~\cite{zhu2024vision}. This versatility has been demonstrated in various visual tasks, including ImageNet classification~\cite{zhu2024vision}, remote sensing image classification~\cite{chen2024rsmamba}, image dehazing~\cite{zheng2024u}, point cloud analysis~\cite{liang2024pointmamba}, and medical image segmentation~\cite{ruan2024vm}, showcasing Mamba's potential beyond driver drowsiness detection and opening new avenues for research in computational neuroscience.

This paper introduces a driver drowsiness detection system using EEG data and the Mamba state-space model. Mamba's ability to handle complex brain activity dynamics makes it ideal for analyzing drowsiness-related EEG changes. The system leverages Mamba's robustness and adaptability to noise and non-linearity in EEG signals. This Mamba-based approach aims to surpass existing methods by providing a more precise and responsive solution, potentially reducing fatigue-related accidents. Additionally, Mamba's advanced feature extraction capabilities offer broader applications in computational neuroscience and BCIs, as demonstrated by its effectiveness in distinguishing cognitive loads. Integrating Mamba into EEG research holds promise for unlocking new discoveries in brain function. The key contributions of this work are as follows:
\begin{itemize}
    \item This research introduces DrowzEE-G-Mamba, a novel deep learning model leveraging the Mamba state-space model for real-time driver drowsiness detection using EEG data. DrowzEE-G-Mamba surpasses existing methods by achieving a peak accuracy of 83.24\% on the SEED-VIG dataset.
    \item DrowzEE-G-Mamba demonstrates exceptional robustness, maintaining high accuracy across varying model complexities. Notably, it achieves a remarkable 83.24\% accuracy even with a minimal 10.1k parameters. This efficiency translates to faster training, lower memory footprint, and easier deployment on resource-constrained devices.
    \item DrowzEE-G-Mamba exhibits a smaller confidence interval compared to other methods, indicating greater consistency in performance. This, coupled with its adaptability across various computational settings, suggests its potential for diverse practical applications beyond driver drowsiness detection, opening doors for real-time brain activity monitoring in other domains.
\end{itemize}

This paper presents a methodical exploration of EEG-based fatigue detection and its potential for enhancing road safety technologies. In Section~\ref{sec_related}, a review of recent literature on driver drowsiness and vigilance is conducted. Section~\ref{sec_method} details the methodology employed in this research. The empirical findings of the study are presented in Section~\ref{sec_result}. Finally, the discussion in Section~\ref{sec_conclusion} extends beyond the results, exploring the broader implications and future directions for this research.


\section{Related Work}
\label{sec_related}

Early research in EEG-based fatigue detection identified biomarkers such as variations in theta and alpha EEG frequency bands~\cite{huang2009tonic}. Deep learning has further transformed EEG analysis, with Convolutional Neural Networks (CNNs) and Recurrent Neural Networks (RNNs) adeptly handling spatial and temporal data~\cite{sheykhivand2022developing}. Hybrid models combining CNNs with RNNs or other techniques enhance feature extraction~\cite{ardabili2024novel}, offering superior accuracy and computational efficiency for real-time applications~\cite{wang2022gradient}. These models manage large, complex datasets without extensive feature engineering, outperforming traditional methods~\cite{wang2023recent}. However, variability in EEG signals across individuals affects model generalization~\cite{kar2010eeg}.

Driver drowsiness detection utilizes physiological (EEG, ECG, EOG)~\cite{khushaba2010driver}, vehicle behavior (steering, lane departure, pedal use)~\cite{dong2010driver}, and behavioral (facial expressions, head position, eye closure)~\cite{bergasa2006real} signals to assess driver state. Physiological methods are accurate but intrusive, while vehicle-based and behavioral methods offer non-intrusive detection but may be less accurate. Recent advancements integrate multiple detection methods (physiological, behavioral, vehicle-based) for improved drowsiness detection accuracy and reliability. Real-world EEG systems face challenges: discomfort from traditional setups, artifact vulnerability, and inter-individual variability requiring personalized models~\cite{fatourechi2007emg}. Future systems should prioritize comfort (dry electrodes, wireless headsets), robust artifact removal, and real-time processing with efficient algorithms. CNNs effectively extract features from EEG signals~\cite{aggarwal2022review}, and Transformers excel at handling time-series data and capturing long-range dependencies in EEG for tasks like mental state classification and seizure detection~\cite{vaswani2017attention,siddhad2024efficacy}.

State space models (SSMs) offer a powerful tool in neuroscience to decipher complex neural dynamics and behaviors. These models describe systems evolving over time, inferring hidden states and underlying processes from observed neural data~\cite{ruan2024vm}. This allows researchers to gain insights into neural activity, dynamics, and behavior. SSMs are particularly useful for decoding neural activity to infer hidden cognitive states, illustrating how neural populations interact and evolve, and linking neural activity with behavior. A prominent application lies in brain-machine interfaces (BMIs). For example, Wu et al.~\cite{wu2006bayesian} used a Kalman filter (an SSM) for real-time motor cortex decoding. Churchland et al.~\cite{churchland2012neural} analyzed motor cortex dynamics with SSMs. Mante et al.~\cite{mante2013context} studied decision-making in the prefrontal cortex using SSMs. Despite these advantages, such as flexibility for diverse data types, hidden state inference, and prior knowledge integration, challenges remain. These include computational intensity, high-quality data requirements, and difficulty interpreting the biological relevance of inferred hidden states. Future research may focus on improving computational methods, integrating multimodal data, and enhancing model interpretability.

While initially limited by computational demands, SSMs have evolved. The Structured State Space Sequence Model (S4)~\cite{gu2021efficiently} addresses this with efficient kernel computations. Additionally, SSMs are now integrated into various deep learning architectures~\cite{smith2022simplified}. However, constant sequence transformation restricts context-based reasoning in standard models. Recent advancements like Mamba (Selective SSM) introduce time-varying parameters for more efficient training and inference~\cite{gu2023mamba}. This paves the way for applying SSMs to computer vision tasks, similar to Transformers in NLP. Studies like ViS4mer~\cite{islam2022long} and S4ND~\cite{nguyen2022s4nd} utilize SSM blocks for modeling visual data across dimensions. VMamba~\cite{liu2024vmamba} and Vim~\cite{zhu2024vision} address direction-sensitivity and global context modeling, respectively. SSMs are a powerful framework in neuroscience, providing deep insights into neural dynamics and behavior. They decode neural activity, model population dynamics, and study cognitive processes. As computational techniques and data quality improve, SSMs are likely to play an even more critical role in advancing our understanding of the brain.


\begin{figure*}[!t]
    \centering
    \includegraphics[width=\textwidth]{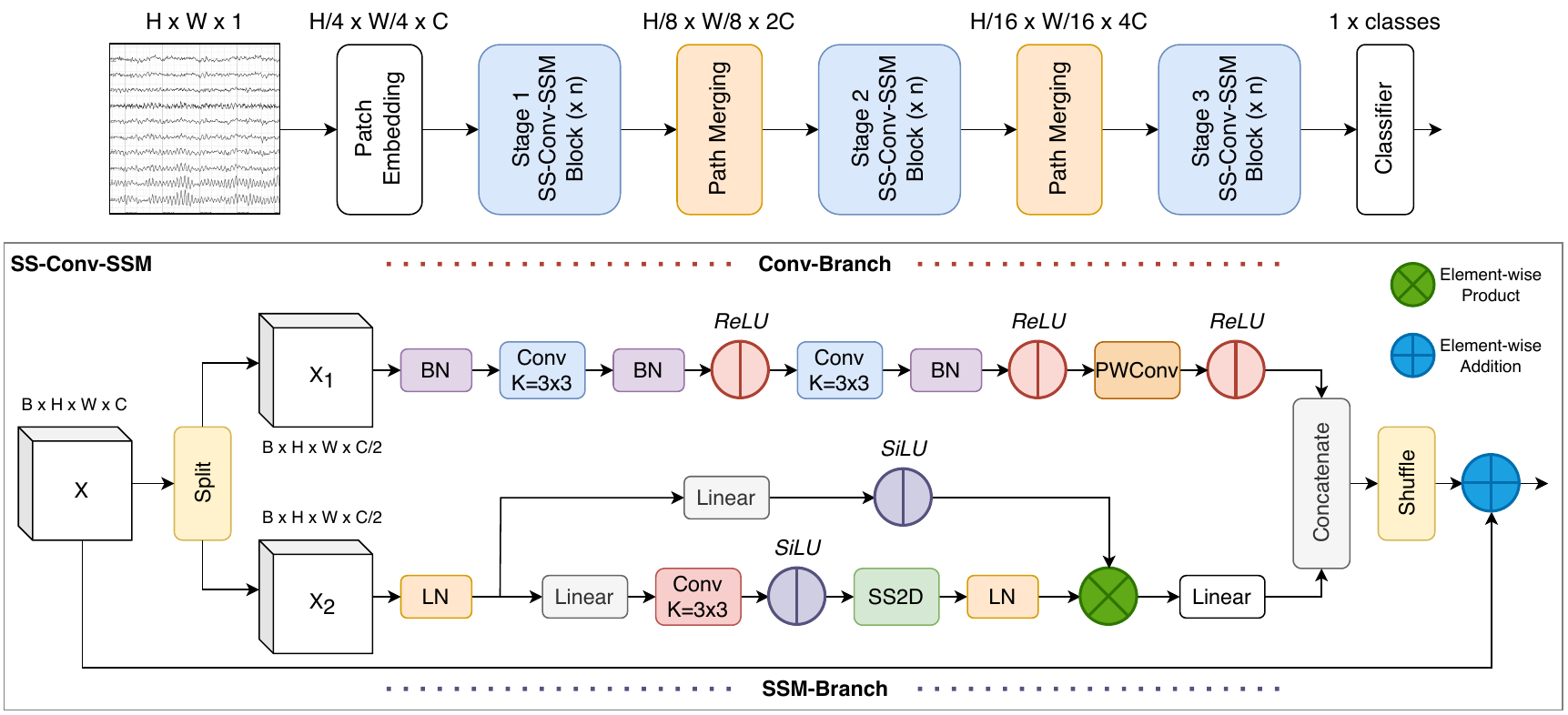}
    \caption{Architecture of DrowzEE-G-Mamba: BN, LN, linear, PWConv, and DWConv represent batch normalization, layer normalization, linear layer, point-wise convolution, and depth-wise convolution, respectively.}
    \label{fig_method}
\end{figure*}

\section{Methodology}
\label{sec_method}

This section examines the foundational concepts underlying DrowzEE-G-Mamba, a deep learning model designed for driver drowsiness detection using EEG data. These concepts, such as State Space Models (SSMs) and their discretization process, are essential for capturing the complex relationships within EEG signals. DrowzEE-G-Mamba's overall architecture is then discussed which is adapted from MedMamba~\cite{yue2024medmamba}. 2D-Selective-Scan mechanism, adapted from VMamba~\cite{liu2024vmamba}, is highlighted as crucial for extracting informative features from the EEG data. Finally, the detailed modeling process of the SS-Conv-SSM block, the fundamental building block of DrowzEE-G-Mamba, is examined to understand how features indicative of drowsiness are efficiently extracted from EEG signals.


\subsection{Preliminaries}
Recent SSM-based models, such as Structured State Space Sequence Models (S4) and Mamba, utilize a classical continuous system to map a 1D input function or sequence, denoted as $x(t) \in \mathcal{R}$, through intermediate implicit states $h(t) \in \mathcal{R}^N$, to an output $y(t) \in \mathcal{R}$. This process can be represented by a linear Ordinary Differential Equation (ODE)~\cite{gu2023mamba,liu2024vmamba}:
\begin{equation}
\begin{aligned}
    &h^{\prime}(t) = \mathbf{A}h(t) + \mathbf{B}x(t) \\
    &y(t) = \mathbf{C}h(t)
\end{aligned}
\label{eq1}
\end{equation} 
Here, $\mathbf{A} \in \mathcal{R}^{N \times N}$ represents the state matrix, while $\mathbf{B} \in  \mathcal{R}^{N \times 1}$ and $\mathbf{C} \in  \mathcal{R}^{N \times 1}$ denote the projection parameters.

The S4 Model and Mamba leverage discretization to make continuous systems compatible with deep learning architectures. This process introduces a timescale parameter, denoted by $\mathbf{\Delta}$, which transforms the continuous system matrices $\mathbf{A}$ and $\mathbf{B}$ into their discrete counterparts, denoted by $\overline{\mathbf{A}}$ and $\overline{\mathbf{B}}$. A common discretization rule employed for this purpose is the zero-order hold (ZOH).
\begin{equation}
\begin{aligned}
    &\overline{\mathbf{A}} = \textup{exp}(\mathbf{\Delta} \mathbf{A}) \\
    &\overline{\mathbf{B}} = (\mathbf{\Delta} \mathbf{A})^{-1}(\textup{exp}(\mathbf{\Delta} \mathbf{A}) - \mathbf{I})\cdot\mathbf{\Delta} \mathbf{B}
\end{aligned}
\label{eq2}
\end{equation}

After applying discretization with a step size $\mathbf{\Delta}$, Equation~\ref{eq1} transforms into a linear recurrence form (Equation~\ref{eq:linear_recurrence}) as follows:
\begin{equation}
\begin{aligned}
    &h^{\prime}(t) = \overline{\mathbf{A}}h(t) + \overline{\mathbf{B}}x(t) \\
    &y(t) = \mathbf{C}h(t)
\end{aligned}
\label{eq:linear_recurrence}
\end{equation}
This equation represents the state update ($h'$) based on the previous state ($h$) and the current input ($x$). Additionally, the output ($y$) is obtained by multiplying the current state with an output matrix ($\mathbf{C}$).

Finally, the SSM model employs a global convolution to efficiently capture long-range dependencies within the input sequence:
\begin{equation}
\begin{aligned}
    &\overline{K} = (\mathbf{C}\overline{\mathbf{B}}, \mathbf{C}\overline{\mathbf{AB}}, \ldots, \mathbf{C}\overline{\mathbf{A}}^{L-1}\overline{\mathbf{B}}) \\
    &y = x * \overline{\mathbf{K}}
\end{aligned}
\label{eq:global_convolution}
\end{equation}
This convolution utilizes a structured kernel ($\overline{\mathbf{K}}$), which incorporates the discretized state transition matrices ($\overline{\mathbf{A}}$, $\overline{\mathbf{B}}$) and the output matrix ($\mathbf{C}$). The length of the input sequence $x$ is denoted by $L$.


\subsection{DrowzEE-G-Mamba Architecture}
DrowzEE-G-Mamba is a deep learning model proposed for driver drowsiness detection. It takes inspiration from the architectural design and concepts of MedMamba and VMamba. It utilizes a patch embedding layer to convert raw EEG data into a format suitable for subsequent processing. The model’s core consists of stacked SS-Conv-SSM blocks, to capture complex spatio-temporal features within EEG signals indicative of drowsiness. Patch merging layers downsample the extracted features, facilitating efficient processing and classification. Finally, a feature classifier accurately identifies drowsiness states based on the learned feature representations.

Figure~\ref{fig_method} illustrates the DrowzEE-G-Mamba model architecture, which processes EEG data in a series of multiple stacked stages. The model begins by transforming the raw EEG data (dimensions $H \times W \times 1$) into a format suitable for subsequent processing through a patch embedding layer. The data then undergoes a series of processing stages, each consisting of multiple SS-Conv-SSM blocks followed by patch merging operations. These stages progressively reduce the spatial dimensions of the feature maps while increasing the channel dimensions. The final output of this processing pipeline is fed into a classifier, which predicts the driver's drowsiness state.

The core of DrowzEE-G-Mamba lies in its stacked SS-Conv-SSM blocks (detailed structure in Figure~\ref{fig_method} bottom section). These blocks are specifically designed to capture the intricate spatio-temporal features within EEG signals that are crucial for drowsiness detection. Each block consists of two branches: Conv-Branch and SSM-Branch. Conv-Branch focuses on extracting local features through standard operations like batch normalization (BN), convolutions (Conv), pointwise convolutions (PWConv), and ReLU activations. SSM-Branch leverages linear layers, depth-wise convolutions (DWConv), SiLU activations, and structured state space 2D (SS2D) components to capture long-range dependencies and global context within the EEG data. Finally, element-wise addition and concatenation operations combine the features from both branches.

A key aspect is the inclusion of a shuffle operation at the end of the block. This helps mitigate potential information loss caused by the initial channel split within the SS-Conv-SSM architecture. This dual-branch design empowers DrowzEE-G-Mamba to efficiently learn complex patterns from EEG data, making it well-suited for driver drowsiness detection and other cognitive state analysis tasks. Inspired by ViTs, DrowzEE-G-Mamba employs a patch embedding layer as the first processing step. This layer transforms the raw EEG data, denoted as $x \in {R^{H \times W \times 1}}$, into non-overlapping patches of size $4 \times 4$. The patch embedding layer achieves this transformation by mapping the single channel dimension to a higher dimensionality ($C$) without flattening the EEG data into a one-dimensional sequence. This approach preserves the two-dimensional (2D) structure of the EEG data, which is crucial for capturing spatial relationships within the signals. As a result, the patch embedding layer generates a feature map with dimensions $\frac{H}{4} \times \frac{W}{4} \times C$.

Following the patch embedding, DrowzEE-G-Mamba leverages stacked SS-Conv-SSM blocks in Stage 1 to process the feature map. These blocks are designed to extract informative features from the EEG data. Crucially, they capture both local details and long-range dependencies within the signals. Importantly, the dimensions of the feature map remain unchanged in this stage, allowing the model to focus on extracting rich features without altering the spatial resolution. To create hierarchical representations of the EEG data, patch merging layers are employed after Stage 1. These layers perform down-sampling, progressively reducing the spatial resolution (denoted by $H$ and $W$) of the feature maps. In contrast, the channel dimension (denoted by $C$) typically doubles after each patch merging layer. Stages 2, 3, and 4 repeat this process, resulting in progressively lower spatial resolutions (e.g., $\frac{H}{16} \times \frac{W}{16} \times 4C$ for Stage 2) and increased channel dimensions. This down-sampling allows the model to learn complex patterns across different scales of the EEG data while maintaining computational efficiency. At the end of the network, a classifier with an adaptive global pooling layer and a linear layer determines the category of the input.


\subsection{2D Selective Scan}
 
The 2D-selective-scan (SS2D) proposed by VMamba, is a core element of MedMamba. SS2D adapts the selective scan space state sequence model (S6) designed for natural language processing to address the ``direction-sensitive'' problem in S6. To bridge the gap between 1-D array scanning and 2-D plane traversing, SS2D introduces a Cross-Scan Module (CSM). CSM uses a four-way scanning strategy, scanning from four corners across the feature map to the opposite locations, ensuring each pixel integrates information from all directions, achieving a global receptive field without increasing computational complexity.

By incorporating CSM, SS2D maintains the linear complexity of S6 while capturing long-range dependencies, essential for accurate medical image classification. SS2D comprises three components: a scan expanding operation (CSM), an S6 block, and a scan merging operation. The scan expanding operation unfolds the input image along four directions (top-left to bottom-right, bottom-right to top-left, top-right to bottom-left, and bottom-left to top-right) into sequences. The S6 block processes these sequences to extract features, ensuring thorough scanning from various directions. Finally, the four directional features are merged through scan merging to reconstruct the 2D feature map, resulting in an output of the same size as the input. The S6 block, derived from Mamba, introduces a selective mechanism based on S4 by adjusting SSM parameters according to input. This enables the model to distinguish and retain relevant information while filtering out irrelevant details. The detailed pseudo-code for the S6 block can be found in the MedMamba~\cite{yue2024medmamba}.


\subsection{SS-Conv-SSM Block}

A hybrid basic block named SS-Conv-SSM, utilized in this work was introduced in MedMamba~\cite{yue2024medmamba}. This block integrates convolutional layers for extracting local features with SSM's ability to capture long-range dependencies. A grouped convolution, introduced in AlexNet~\cite{krizhevsky2017imagenet}, uses multiple kernels per layer to promote learning various high and low level features was also incorporated into the SS-Conv-SSM. SS-Conv-SSM is a lightweight dual-branch block (Figure~\ref{fig_method}). It partitions the feature map into two groups using channel-split, then extracts global and local information from each group through the Conv-Branch and SSM-Branch, respectively. Finally, channel-concatenation restores the channel dimension size, and channel-shuffle ensures information is not lost between channels due to grouped convolution operations~\cite{zhang2018shufflenet}. Following the settings of classic CNNs and ViTs, the activation functions in the Conv-Branch and SSM-Branch are set to ReLU~\cite{agarap2018deep} and SiLU~\cite{elfwing2018sigmoid}, respectively.

The modeling process of SS-Conv-SSM for feature maps is formalized. Given a module input $x \in {R^{H \times W \times C}}$ and a module output $y \in {R^{H \times W \times C}}$, $f$ is used to represent the channel-split, and then there is \[x \in {R^{H \times W \times C}}{x_{i = 1,2}} \in {R^{H \times W \times \frac{C}{2}}}\] 

Next, the ${f^{ - 1}}$ and $g$ are used to represent channel-concatenation and channel-shuffle respectively. To match the convolution operation, a permute operation is utilized to rearrange the original feature map. Based on the above, the modeling process of Conv-Branch can be defined as follows:
\[\overline {{x_1}}  \in {R^{\frac{C}{2} \times H \times W}} \leftarrow permute({x_1})\]
\[{x_1}^\prime  = BatchNor{m_1}(\overline {{x_1}} )\]
\[{x_1}^{\prime \prime } = ReLU(BatchNor{m_2}(Con{v_{3 \times 3}}({x_1}^\prime )))\]
\[{x_1}^{\prime \prime \prime } = ReLU(BatchNor{m_3}(Con{v_{3 \times 3}}({x_1}^{\prime \prime })))\]
\[\widehat {{x_1}} = ReLU(PWConv({x_1}^{\prime \prime \prime }))\]
\[\widetilde {{x_1}} \in {R^{H \times W \times \frac{C}{2}}} \leftarrow permute(\widehat {{x_1}})\]

Meanwhile, the modeling process of SSM-Branch can be defined as follows:
\[\overline {{x_2}}  = LayerNor{m_1}({x_2})\]
\[{x_2}^\prime  = SiLU(DWConv(Linear(\overline {{x_2}} )))\]
\[{x_2}^{\prime \prime } = LayerNor{m_2}(SS2D({x_2}^\prime ))\]
\[{x_2}^{\prime \prime \prime } = SiLU(Linear(\overline {{x_2}} ))\]
\[\widetilde {{x_2}} = Linear({x_2}^{\prime \prime } \otimes {x_2}^{\prime \prime \prime })\]

In summary, the output of SS-Conv-SSM be formulated as follows:
\[y = x \oplus g({f^{ - 1}}(\widetilde {{x_1}},\widetilde {{x_2}}))\]


\section{Results and Discussion}
\label{sec_result}

This section presents the findings of this study and analyzes their significance for the field of driver drowsiness research. The analysis focuses on the effectiveness of the employed methods and the implications of the observed outcomes. This is followed by a comparative analysis with relevant findings from existing literature to contextualize our results.


\subsection{Experimental Data}

This study utilizes the SEED-VIG dataset~\cite{zheng2017multimodal}, a valuable open-source resource designed to investigate driver vigilance and drowsiness through EEG recordings. The dataset offers a diverse subject pool, encompassing recordings from 23 participants. To enhance real-world applicability, participants engaged in a driving simulation designed to closely mimic real-world driving conditions. EEG recordings were captured using a 17-channel montage based on the international 10-20 system. This montage specifically targeted key temporal (FT7, FT8, T7, T8, TP7, TP8) and posterior (CP1, CP2, P1, PZ, P2, PO3, POZ, PO4, O1, OZ, O2) brain regions, ensuring comprehensive coverage of brain activity relevant to vigilance and drowsiness. High temporal resolution, crucial for detailed analysis, was achieved with a sampling rate of 1000 Hz. Sessions were strategically scheduled post-lunch to encourage the onset of fatigue in participants.

Drowsiness states were quantified using the PERCLOS (percentage of eyelid closure) metric. A threshold of 0.5 was employed to classify PERCLOS values into ``awake'' and ``drowsy'' states, enabling a binary classification approach for evaluating driver fatigue detection methods. To minimize artifacts and improve computational efficiency, EEG signals were band-pass filtered (1-75 Hz) and down-sampled to 200 Hz. Subsequently, the data was segmented into one-second epochs, resulting in a standardized format of (17, 200, 1) per epoch. The entire dataset comprised approximately 40,710 samples and was divided into training (70\%), validation (15\%), and test (15\%) sets to facilitate model development and evaluation.


\subsection{Implementation Details}

The computational environment consisted of a DELL Precision 7820 Tower Workstation equipped with Ubuntu 22.04 operating system, an Intel Core(TM) Xeon Silver 4216 CPU, and an NVIDIA RTX A4000 12GB GPU. This hardware configuration facilitated the implementation of Deep Learning (DL) models using Python 3.12 and the PyTorch library. The Adam optimizer, recognized for its efficiency, was employed with its default hyperparameters ($\eta$ = 0.001, $\beta_1$ = 0.9, $\beta_2$ = 0.999). Both EEGNet and TSception models underwent training for 100 epochs, utilizing a batch size of 16 and a learning rate of $1e-4$. For the Support Vector Machine (SVM) classification, the Radial Basis Function (RBF) kernel from scikit-learn~\cite{sklearn} was implemented with its default settings. Stratified five-fold cross-validation was employed to assess classification accuracy, with the results averaged for a robust evaluation.


\subsection{Classifiers}

This work employs a balanced evaluation approach using three established classifiers for EEG-based emotion classification. Support Vector Machine (SVM)~\cite{cortes1995support} is a popular supervised learning model for classification, known for its ability to maximize the class margin for new data points. SVMs can handle non-linear classification through the kernel trick, effectively mapping inputs into high-dimensional spaces. EEGNet~\cite{lawhern2018eegnet} is a CNN-based architecture that achieves competitive accuracy using deep and separable convolutions. It incorporates temporal convolution for learning frequency filters, depth-wise convolution for frequency-specific spatial filters, and separable convolution for efficient feature map combinations. TSception~\cite{ding2022tsception} utilizes a dynamic temporal layer to learn temporal and frequency representations from EEG channels. It also includes an asymmetric spatial layer for capturing global spatial patterns and emotional asymmetry, a high-level fusion layer, and a final classifier that leverages various convolutional kernel sizes for spatial analysis. ConvNext~\cite{liu2022convnet} is a state-of-the-art CNN architecture that achieves competitive performance on various image classification benchmarks. It incorporates design principles from recent transformer models to enhance feature learning and improve efficiency compared to traditional CNNs. LMDA-Net~\cite{miao2023lmda} is a lightweight deep learning model specifically designed for EEG-based emotion classification. It employs a multi-modal approach, combining temporal and spatial features, to effectively capture the complex patterns in EEG signals, resulting in efficient and accurate emotion recognition.


\subsection{Evaluation}

\begin{table}[!t]
    \centering
    \caption{Results of different methods on SEED-VIG dataset for driver drowsiness detection with 95\% confidence interval}
    \label{tab_results}
    \begin{tabular}{l c}
        \toprule
        \textbf{Method} & \textbf{Accuracy}\\
        \midrule
        
        \textbf{SVM}~\cite{cortes1995support} & $65.52 \pm 0.02$ \\
        \textbf{EEGNet}~\cite{lawhern2018eegnet} & $80.74 \pm 0.75$ \\
        \textbf{TSception}~\cite{ding2022tsception} & $83.15 \pm 0.36$ \\
        \textbf{ConvNeXt}~\cite{liu2022convnet} & $81.95 \pm 0.61$ \\
        \textbf{LMDA-Net}~\cite{miao2023lmda} & $81.06 \pm 0.99$ \\

        \midrule
        \textbf{DrowzEE-G-Mamba} & $\mathbf{83.24 \pm 0.24}$ \\
        \bottomrule
    \end{tabular}
\end{table}

The results presented in Table~\ref{tab_results} demonstrate the effectiveness of different methods for driver drowsiness detection on the SEED-VIG dataset. The evaluation revealed a clear hierarchy in the effectiveness of the compared methods for driver drowsiness detection on the SEED-VIG dataset. Support Vector Machine (SVM) achieved the lowest accuracy (65.52\%) with a narrow confidence interval (0.02), indicating consistent but limited performance. This suggests SVM may not adequately capture the complexities of EEG data for this task. EEGNet demonstrated a significant improvement over SVM, achieving an accuracy of 80.74\%. However, its larger confidence interval (0.75) implies greater variability in performance. While superior to SVM, this suggests EEGNet might benefit from further optimization for drowsiness detection. TSception surpassed EEGNet with an accuracy of 83.15\% and a reduced confidence interval (0.36), indicating both higher accuracy and more consistent performance. This suggests TSception's architecture effectively captures relevant features in the EEG data. ConvNeXt achieved an accuracy of 81.95\% and LMDA-Net obtained an accuracy of 81.06\%. While their performance was comparable, DrowzEE-G-Mamba's higher accuracy and lower variability make it a more reliable choice for real-time driver drowsiness detection.

DrowzEE-G-Mamba emerged as the most effective method, achieving the highest accuracy (83.24\%) with the smallest confidence interval (0.24). This signifies not only superior detection accuracy but also the most consistent results. By combining EEG data with State Space Models (SSMs), DrowzEE-G-Mamba effectively models both local and long-range dependencies within the data, leading to superior drowsiness state detection. In conclusion, these findings highlight the clear advantage of DrowzEE-G-Mamba compared to traditional methods (SVM) and advanced neural network approaches (EEGNet, TSception, ConvNeXt, and LMDA-Net) for driver drowsiness detection on the SEED-VIG dataset. Its high accuracy and low variability make DrowzEE-G-Mamba a promising tool for real-time driver drowsiness detection, potentially contributing to accident prevention and improved road safety.
 
\begin{figure}[!t]
    \centering
    \includegraphics[width=\linewidth]{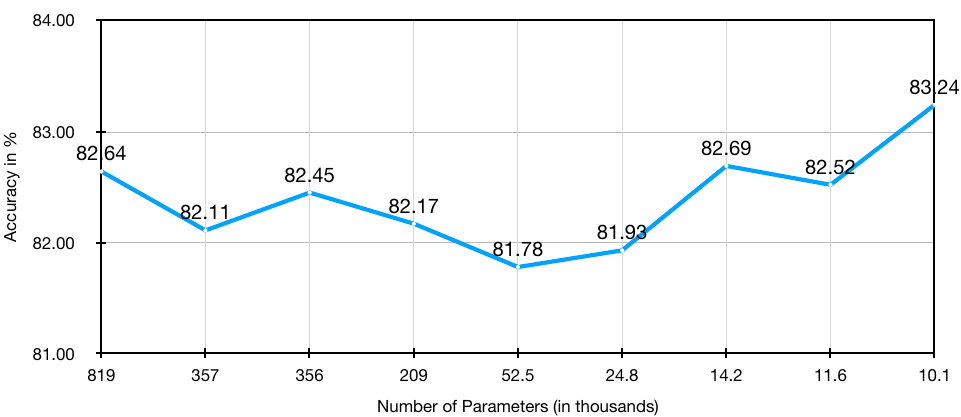}
    \caption{This chart shows the impact of model complexity on driver drowsiness detection using the SEED-ViG dataset. It visualizes the relationship between average model accuracy (percentage) with 95\% confidence interval and the number of parameters (thousands). As evident, accuracy increases with model complexity, ranging from 81.78\% for an 819K parameter model to 83.24\% for a 10.1K parameter model. One of the primary strategies employed to reduce the model size was the careful adjustment of hyperparameters, specifically through the elimination of certain blocks within the model architecture.}
    \label{fig_ablation}
\end{figure}


The chart in Figure~\ref{fig_ablation} illustrates the accuracy of the DrowzEE-G-Mamba model on the SEED-VIG dataset for driver drowsiness detection, plotted against the number of parameters (in thousands). The model achieves an accuracy of 82.64\% with 819k parameters. As the number of parameters decreases, the accuracy generally remains above 81\%, with slight fluctuations. For instance, at 357k parameters, the accuracy is 82.11\%, while at 209k parameters, it is 82.17\%. The lowest number of parameters tested is 10.1k, where the accuracy maintains a robust peak at 83.24\%. This chart demonstrates that DrowzEE-G-Mamba maintains high accuracy across a range of model complexities, with only a minor fluctuation in performance as the number of parameters decreases. The model which gave the highest accuracy with 10.1k parameters had one SS-Conv-SSM block with 32 dimensions. 

The combined analysis of the presented table and chart suggests DrowzEE-G-Mamba's exceptional potential as a highly effective and reliable model for driver drowsiness detection. Notably, the model achieves accuracy levels higher than the peak performance of leading models like TSception. Furthermore, DrowzEE-G-Mamba demonstrates a remarkable characteristic, it maintains this high accuracy across varying levels of model complexity (as reflected by different parameter counts). This robustness makes DrowzEE-G-Mamba particularly well-suited for real-time applications where computational resources might be constrained. The model's consistency in performance is further emphasized by the narrow confidence interval and stable accuracy observed across parameter counts. This consistency underscores DrowzEE-G-Mamba's suitability for practical deployment in real-world scenarios.


\section{Conclusion}
\label{sec_conclusion}

This research investigated the efficacy of DrowzEE-G-Mamba, a deep learning model for driver drowsiness detection using EEG data. DrowzEE-G-Mamba achieved a peak accuracy of 83.24\% on the SEED-VIG dataset, demonstrating its effectiveness. Notably, the model maintained high accuracy across varying parameter complexities, indicating strong robustness for real-time applications with limited computational resources. DrowzEE-G-Mamba's architecture balances sophistication with efficiency. The model leverages channel-split, channel-concatenation, and channel-shuffle operations to optimize information flow within the EEG data. DrowzEE-G-Mamba surpasses existing methods in two key aspects: accuracy and robustness. It achieves the highest accuracy while maintaining this performance even with a significant number of parameters. This translates to consistent and reliable detection, even with a larger computational footprint, making it a strong candidate for real-time driver drowsiness detection.

Overall, DrowzEE-G-Mamba presents a robust, efficient, and highly accurate solution for driver drowsiness detection. Its ability to function across diverse computational constraints makes it a promising tool for real-time drowsiness monitoring and enhancing road safety. Future work will focus on further optimization and explore applications in broader cognitive state detection tasks, expanding its impact and utility in various real-world scenarios. While challenges remain in refining accuracy and generalizability of fatigue detection systems, DrowzEE-G-Mamba's performance highlights the potential for significant advancements in real-time driver fatigue detection. Future research will target further accuracy improvements, applicability expansion, and integration into practical, real-world applications, ultimately contributing to safer driving environments.




\section*{Bibliography}
\bibliography{references}








\end{document}